\documentclass[aps,prl,twocolumn,floatfix,showpacs,showkeys,amssymb]{revtex4}

\usepackage{graphicx}

\usepackage{amsmath}

\usepackage{amsfonts}

\usepackage{amssymb}

\begin{document}
\topmargin -15mm

\def\ba{\begin{array}}

\def\ea{\end{array}}

\def\be{\begin{equation}\begin{array}{l}}

\def\ee{\end{array}\end{equation}}

\def\bea{\begin{equation}\begin{array}{l}}

\def\eea{\end{array}\end{equation}}

\def\f#1#2{\frac{\displaystyle #1}{\displaystyle #2}}

\def\om{\omega}

\def\Om{\Omega}

\def\omm{\omega^a_b}

\def\we{\wedge}

\def\de{\delta}

\def\De{\Delta}

\def\va{\varepsilon}

\def\omb{\bar{\omega}}

\def\la{\lambda}

\def\vv{\f{V}{\la^d}}

\def\si{\sigma}

\def\t{T_+}

\def\v{v_{cl}}

\def\m{m_{cl}}

\def\n{N_{cl}}

\def\bi{\bibitem}

\def\c{\cite}

\def\sa{\sigma_{\alpha}}

\def\ua{\uparrow}

\def\da{\downarrow}

\def\mua{\mu_{\alpha}}

\def\ga{\gamma_{\alpha}}

\def\g{\gamma}

\def\ora{\overrightarrow}

\def\pa{\partial}

\def\ov{\ora{v}}

\def\al{\alpha}

\def\bt{\beta}

\def\R{R_{\rm eff}}

\def\muu{\f{\mu}{ed}}

\def\E{\f{edE(\tau)}{\om}}

\def\t{\tau}

\title{Semiconductor-metal nanoparticle molecules: hybrid excitons and
non-linear Fano effect} 

\author{Wei Zhang $^1$, Alexander O. Govorov $^1$ and Garnett W. Bryant
$^2$}

\affiliation{$~^1$ Department of Physics and Astronomy,
Ohio University, Athens, OH 45701-2979 \\
$~^2$ National Institute of Standards and Technology,
Gaithersburg, MD 20899-8423 }
\date{\today } 

\begin{abstract}



Modern nanotechnology opens the possibility of combining
nanocrystals of various materials with very different
characteristics in one superstructure. The resultant
superstructure may provide new physical properties not encountered
in homogeneous systems.   Here we study theoretically the optical
properties of hybrid molecules composed of semiconductor and metal
nanoparticles. Excitons and plasmons in such a hybrid molecule
become strongly coupled and demonstrate novel properties. At low
incident light intensity, the exciton peak in the absorption
spectrum is broadened and shifted due to incoherent and
coherent interactions between metal and semiconductor
nanoparticles. At high light intensity, the absorption spectrum
demonstrates a surprising, strongly asymmetric
shape.   This shape originates from the coherent
inter-nanoparticle Coulomb interaction and can be viewed as
a non-linear Fano effect which is quite different from the
usual linear Fano resonance.
\end{abstract}



\maketitle




Modern nanoscience involves both solid-state nanostructures
and bio-materials. Using bio-molecules as linkers, solid-state
nanocrystals with modified surfaces can be assembled into
superstructures. Building blocks of these superstructures are
nanowires, semiconductor quantum dots (SQD), metal nanoparticles
(MNP), proteins, etc.\c{bionano,etrans}. It is important to
emphasize that these building blocks are composed of different
materials. Thus, if the building blocks interact, the resulting
superstructure may have unique physical properties. To date,
several interesting phenomena in bio-conjugated colloidal
nanocrystals, such as energy transfer \c{etrans}, local field
enhancement \c{eenhan}, and thermal effects
\c{Lee_2005_AC,Richardson_2006_NL}, have been explored. In
parallel with bio-assembly, self-organized growth of epitaxial
SQDs has become well developed. Self-assembled SQDs have
excellent optical quality and atomically-sharp optical lines
\c{Gammon-Alex_H-Karrai}. Furthermore, using special growth
techniques, SQDs can be arranged in 1D and 2D
structures\c{Salamo}. This capability for nano/bio/epitaxial assembly
opens up the fabrication of complex hybrid superstructures that could
exploit the discrete optical response of excitons in semiconductor
nanosystems and the strong optical response of plasmons in metallic
nanoparticles.

In this paper, we reveal novel unusual optical properties that arise in
hybrid MNP-SQD molecules and motivate experiments to investigate these properties.
Hybrid molecules have already been bio-assembled and studied at room
temperature ($T$) by several groups \c{etrans, eenhan,
Lee_2005_AC}. Hybrid MNP-SQD complexes can also potentially be
realized with epitaxial SQDs. First attempts to integrate MNP
into a solid matrix of GaAs was done in the work \c{Gossard}. In
the future, such MNPs built into a crystal matrix could be combined
with epitaxial SQDs to realize hybrid nanocrystal
molecules. The experimental methods to study such nanostructures
include photoluminescence and absorption spectroscopies
\c{Gammon-Alex_H-Karrai}, and Raleigh scattering. While the effect
of energy transfer from SQD to MNPs can be observed at room $T$,
the fine-scale coherent effects of inter-nanocrystal interaction can
become accessible only at low $T$. For such studies, nanocrystal
molecules should be deposited on surface or embedded into a
polymer or solid matrix. In the
current state-of-art spectroscopy of epitaxial SQDs at low $T$, the
exciton lines may have a $\mu eV$ width \c{Gammon-Alex_H-Karrai}.
A combination of lithography, epitaxial growth, and
bio-assembly could be used to fabricate 3D structures with desired
geometry.

In this letter, we study the optical properties of hybrid
structure composed of a MNP and a SQD. We explore both the linear
regime (for weak external field) and the non-linear regime (for
strong external field). The basic excitations in the MNP are the
surface plasmons with a continuous spectrum. In SQDs, the
excitations are the discrete interband excitons. In the hybrid
structure there is no direct tunnelling between the MNP and the
SQD. However, long-range Coulomb interaction couples the excitons
and plasmons and leads to the formation of hybrid excitons and to
F\"orster energy transfer. The effect of coupling
between excitons and plasmons becomes especially strong near
resonance when the exciton energy lies in the vicinity of
the plasmon peak. The coupling between the continuum excitations
and the discrete excitations also leads to a novel effect that we
call a non-linear Fano effect. We should note that the usual Fano
effect was introduced for the linear regime \c{fano}. Here we
describe a non-linear Fano effect which appears at high intensity
of light when the SQD becomes strongly excited. This nonlinear
Fano effect comes from interference between the external field and
the induced internal field in the hybrid molecule. It appears at
high intensities when the degree of coherence in the system
becomes strongly increased because the Rabi frequency starts to
exceed the exciton broadening.

We now consider a hybrid molecule composed of a spherical MNP of radius $a$
and a spherical SQD with radius $r$ in the presence of polarized external field
$E=E_0cos(\om t)$, where the direction of polarization is specified below.
The center-to-center distance
between the two nanoparticles is $R$ (see the insert of Fig.~1). For
the description of the MNP we use classical electrodynamics and
the quasistatic approach. Because the hybrid structure is much
smaller (i.e. tens of nanometers) than the wavelength of the
incident light, we can neglect retardation effects. For the SQD
we employ the density matrix formalism and the following model for a spherical
SQD. Due to its symmetry, a spherical SQD has three bright
excitons with optical dipoles parallel to the direction $\alpha$,
where $\alpha$ can be x, y, and z \c{Govorov_NL_2006}. The
dark exciton states are not directly involved because they are not excited in
the dipole limit. However, the dark states do provide a nonradiative decay channel
for the bright excitons which contributes to the exciton lifetime.
Using the symmetry of the molecule and a
linearly polarized internal field we can obtain the appropriate
Hamiltonian \c{Yariv}

\be \hat{H}_{SQD}=\sum_{i=1,2}\epsilon_i
c_i^+c_i-\mu E_{SQD}(c_1^+c_2+c_2^+c_1), \ee
here $c_1^+, c_2^+$
are the creation operators for the vacuum ground state and
$\alpha$-exciton state, respectively; $\mu$ is the interband dipole
matrix element, $E_{SQD}$ is the total field felt by the SQD
\be E_{SQD}=E+\f{s_\alpha P_{MNP}}{\va_{eff1} R^3}, \ee with
$\va_{eff1}=\f{2\va_0 +\va_s}{3\va_0}$, $\va_0$ and $\va_s$ are
the dielectric constants of the background medium and SQD,
respectively; $E$ the external field, $s_\alpha= 2 (-1)$ for
electric field polarizations $\alpha=z (y, x)$. The z-direction
corresponds to the axis of the hybrid molecule. The dipole
$P_{MNP}$ comes from the charge induced on the surface of MNP. It
depends on the total electric field which is the superposition of
external field and the dipole field due to the SQD, \be P_{MNP}=\g
a^3 (E+\f{s_\alpha P_{SQD}}{\va_{eff2} R^3}), \ee where
$\g=\f{\va_m(\omega)-\va_0}{2\va_0+\va_m(\omega)}$,
$\va_{eff2}=\f{2\va_0 +\va_s}{3}$, $\va_m(\omega)$ is the
dielectric constant of the metal. The dipole of the SQD is
expressed via the off-diagonal elements of the density matrix:
$P_{SQD}=\mu (\rho_{21}+\rho_{12})$ \c{Yariv}. These matrix
elements should be found from the master equation: \be
\f{d\rho}{dt}=\f{i}{\hbar} [\rho,H_{SQD}]-\Gamma \rho, \ee where
the diagonal and off-diagonal relaxation matrix elements are
$\Gamma_{12}=\Gamma_{21}=1/T_{20}$ and
$\Gamma_{22}=-\Gamma_{11}=1/ \tau_0$ \c{Yariv}. Here, $\tau_0$
includes the nonradiative decay via the dark states. We note that the
above procedure treats the inter-nanoparticle interaction in the
self-consistent way. To solve the coupled system, we separate the
high frequency part and write $\rho_{12}$ and $\rho_{21}$ as
$\rho_{12}=\bar{\rho}_{12}e^{i\om t}$ and
$\rho_{21}=\bar{\rho}_{21}e^{-i\om t}$. Applying the rotating wave
approximation, we obtain the equations for steady state. Let
$\bar{\rho}_{21}=A+Bi$ and $\De=\rho_{11}-\rho_{22}$, we come to
the system of nonlinear equations \bea
A=-\f{(\Om_I+K\Om_R)T_2}{1+K^2}\De\\
B=\f{(\Om_R-K\Om_I)T_2}{1+K^2}\De\\
(1-\De)/\tau_0=4\Om_R B-4\Om_I A-4G_I (A^2+B^2), \eea where
$K=[(\om-\om_0)+G_R\De]T_2$, $\om_0=(\va_2-\va_1)/\hbar$,
$1/T_2=1/T_{20}+G_I$, $G=\f{s_\alpha^2 \g a^3 \mu^2}{\hbar
\va_{eff1}\va_{eff2} R^6}$, $G_R=Re[G]$, $G_I=Im[G]$,
$\Om_{eff}=\Om_0 [1+s_\alpha \g (\f{a}{R})^3]$, $\Om_0 =\f{\mu
E_0}{2\hbar\va_{eff1}}$, and
$\Om_R=Re[\Om_{eff}],\Om_I=Im[\Om_{eff}]$. Here $\Om_{eff}$ is the
Rabi frequency of the SQD renormalized due to the dipole interaction
with the plasmon of the MNP.

For a weak external field, we obtain the following steady state
solution in the analytical form

\bea
\bar{\rho}_{12}=-\f{\Om_{eff}}{(\om-\om_0+G_R)-i(\Gamma_{12}+G_I)} .
\eea
The plasmon-exciton interaction leads to the formation of a hybrid
exciton with shifted exciton frequency and decreased lifetime
determined by $G_R$ and $G_I$, respectively.
In other words, the long-range Coulomb interaction leads to
incoherent energy transfer via the F\"orster mechanism
with energy transfer rate $G_I$.
The exciton shift  $G_R$ shows that the interaction
is partially coherent.
A similar theoretical formalism
for F\"orster transfer was successfully used to describe
available experiential data \c{Govorov_NL_2006,Lee_2005_AC}.

{\bf Energy absorption.}  The energy absorption rate is
$Q=Q_{MNP}+Q_{SQD}$, where the rate of absorption in the MNP and
SQD are $Q_{MNP}=<\int {\bf j E} dV>$, where ${\bf j}$ is the
current, $<...>$ is the average over time, and $Q_{SQD}=\hbar
\om_0 \rho_{22}/\tau_0 $. As an example, we consider a Au MNP with
radius $a=$7.5 nm. We use the  bulk dielectric constant of Au
$\va_m(\om)$ taken from \c{table}, $\va_0=1$, and $\va_s=6.0$. The
bare exciton frequency $\om_0$ is chosen to be 2.5 eV, close to
the surface plasmon resonance of Au MNP. Typically, the plasmon
peak is very broad compared with the bare exciton peak, thus a
small detuning of the frequencies should have no important effect.
 Both the plasmon resonant frequency and the bare exciton frequency
can be tuned in a wide range of energies (from blue to red) by
changing the size and composition of SQD/MNP. For the relaxation
times and dipole moment, we take $\tau_0=0.8$ ns, $T_{20}=0.3$ ns, and
$\mu=er_0$ and $r_0=0.65$ nm.

\begin{figure}[tbh]
\includegraphics*[width=0.7\linewidth]{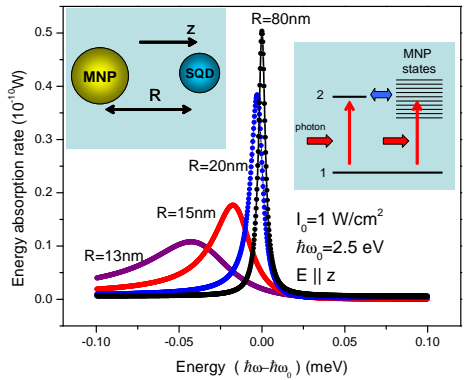}
\caption{ Energy absorption spectra in the weak
field regime for different interparticle distances. The light
intensity is $1$W/cm$^2$. $\om$ is the light frequency. $\hbar \om_0$ is the bare
exciton energy. The left insert shows a model. Right insert: Quantum transitions 
in the system; the vertical (horizontal) arrows represent light
(Coulomb)-induced transitions.
 } \label{fig1}
\end{figure}

In Fig.1, we show the total energy absorption rate versus frequency.
For weak incident light ($I_0=1~W/$cm$^2$), the energy
absorption peak shifts and broadens for
small inter-particle separations $R$. This behavior of the optical
spectrum for relatively small $R$ reflects the formation of the
hybrid exciton with a shifted frequency and shortened life
time. In current experiments on
epitaxial SQDs, the width of the exciton peak can be as small as a few
$\mu$eV \c{Gammon-Alex_H-Karrai}. In our results in Fig.~1, the
frequency shift is about $40~\mu$eV for small separations
$R\approx15$~nm.

\begin{figure}[tbh]
\includegraphics*[width=0.7\linewidth]{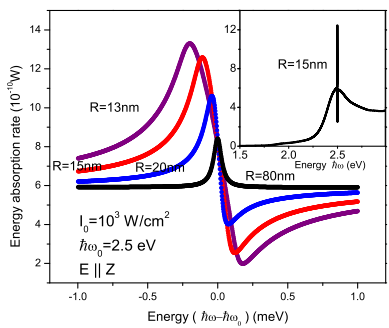}
\caption{Energy absorption spectra in the strong
field regime for different interparticle distances. The insert is
the energy absorption for $R=15$nm for a wider frequency regime; the
exciton feature is within the plasmon peak.} \label{fig2}
\end{figure}

Figure 2 shows the energy absorption in the strong field regime.
We find an asymmetrical Fano shape and substantial suppression of
energy absorption. This striking asymmetry originates from the
Coulomb coupling and vanishes at large $R$ (see Fig. 2).

In the usual linear Fano effect, the absorption intensity becomes
zero for a particular frequency due to the interference effect
\c{fano}. Here we find an nonvanishing energy absorption at any
light frequency. This due to the non-linear nature of the
interference effect in the hybrid molecule. More qualitative
discussion will be provided later. Again we see a red shift of the
resonant frequency. The shift is now one order of magnitude larger
than the energy resolution limit.

\begin{figure}[tbh]
\includegraphics*[width=0.7\linewidth]{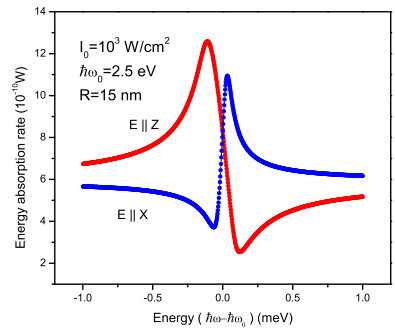}
\caption{Polarization dependence for the energy absorption.}
\label{fig3}
\end{figure}

In Fig.~3, we show the polarization dependence. The Fano
absorption intensity has the opposite shape for the electric field
polarizations along the $z$ and $x(y)$ directions.
The shape reversal due to polarization happens in a frequency
window of 1~meV, and should be an observable experimental signature.

From Eqs. (3) and (5), we see that the effective field
applied to MNP and SQD is the superposition of the external field
and the induced internal field. The interference between external
field and internal field leads to the asymmetric Fano shape. And
enhancement or suppression of the effective field depends on the
polarization ($s$ changes sign for the polarizations $z$ and
$x(y)$). So, the polarization dependence is also a result of
interference of the external field and induced internal field.


{\bf Rayleigh scattering.}  We use the standard method to
calculate Rayleigh scattering intensity \c{jackson}, which is
valid when the size of the scattering objects is much smaller than
the wavelength of incident light: $dI/d\Om=P_0 \sin^2(\theta)$,
where the angle $\theta$ is measured from the direction of the
induced dipole and $P_0=(ck^4/\hbar)(P_{SQD}+P_{MNP})^2$.

\begin{figure}[tbh]
\includegraphics*[width=0.7\linewidth]{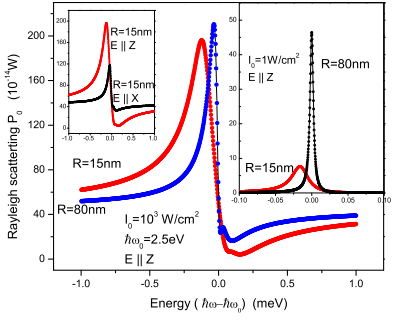}
\caption{Rayleigh scattering intensity $P_0$. The main panel is for
the strong field regime. Right insert:  Scattering intensity $P_0$
for the weak field regime. Left insert: Intensity $P_0$ for
different polarizations in the strong field regime.} \label{fig4}
\end{figure}

Figure 4 shows the Rayleigh scattering in the linear and nonlinear
regimes. Energy absorption and Rayleigh scattering show similar
features, except that the Fano resonance in the Rayleigh
scattering does not reverse its shape when the polarization is
changed. From the above definition of $P_0$, we see that even in the
absence of interaction between MNP and SQD, there is a cross term
in $|P|^2$, which leads to the asymmetric shape of Rayleigh
scattering in the strong field limit. The interaction between SQD
and MNP leads to a shift of the peak of scattering intensity in
Fig.~4 and to the polarization dependence (see insert of Fig. 4).
In fact, 
the interaction effect is sensitive to polarization as we
discussed before, but this dependence on the polarization
may be masked by the asymmetric shape in the absence of
interaction. Actually, the difference between  Rayleigh
scattering for the case with interaction and that for the case
without interaction is sensitive to the polarization
(not shown here).

As discussed above, experiments could be
performed on self-assembled SQDs coupled to MNPs. The insert of Fig.~5
shows a schematic of such a system with $\va_0=\va_s=12$ with the MNP
embedded in the barrier material that defines the SQD (here we still assume
a spherical SQD for simplicity). Again we
see a clear asymmetry which is strong even in the linear regime.
This behavior was found for parameters typical of a
self-assembled SQD. The electric field polarization was chosen
along $y(x)$ directions because self-assembled lens-shape SQDs
have typically two excitons with optical dipoles perpendicular to
the growth direction $z$ (insert of Fig.~5). The strong advantage
of self-assembled SQDs is that an exciton in such systems exhibits
a very narrow line \c{Gammon-Alex_H-Karrai}.

\begin{figure}[tbh]
\includegraphics*[width=0.7\linewidth]{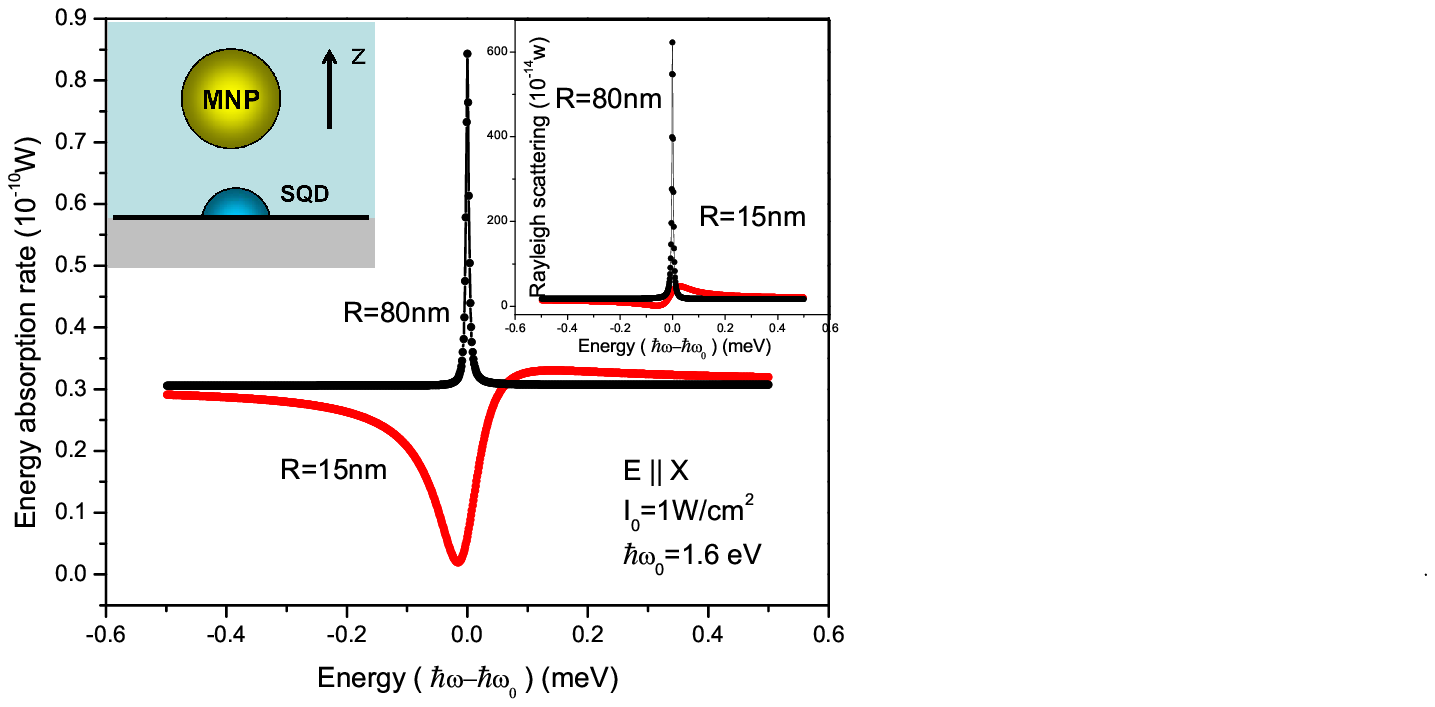}
\caption{Energy absorption and Rayleigh scattering (right insert)
for a system with a self-assembled quantum dot in the weak field
regime. The left insert: schematics of hybrid molecule. The
self-assembled SQD is formed at the interface of two materials. }
\label{fig5}
\end{figure}


{\bf Nonlinear Fano effect.}  In the linear regime, the absorption peak can
have almost a Lorentzian shape (Fig.~1). Here we describe a non-linear Fano effect.
This effect manifests
itself as a strong asymmetry of the absorption peak at high light
intensities.

The energy absorption in the weak field regime/strong field regime
is
\be Q = C\Om_{0}^2 (\f{(K-q)^2}{1+K^2}+\alpha \f{1}{1+K^2})+ \beta
\f{ \hbar \om_0}{1+K^2}. \ee
Here $K=[(\om-\om_0)+G_R\De]T_2$, $q=\f{s_\alpha \mu^2 \Om_R T_2
\De}{\hbar \va_{eff1}\va_{eff2} R^3 \Om_0}$ and the coefficient
$C=\f{1}{6}(\f{2\hbar}{\mu})^2a^3 \om
|\f{3\va_0}{2\va_0+\va_m}|^2Im[\va_m]$. For the weak field regime,
$\alpha=(\f{1}{1+G_I T_{20}})^2$ and $\beta=|4\Om_{eff}|^2
\f{T_2^2}{T_{20}}$, and for the strong field regime $\alpha=1$ and
$\beta=1/2\tau_0$.

In the weak field regime ($\Om_{0}\ll 1/T_2,1/\tau$),
$Q$ has the Fano function form in the
limit $T_{20} \rightarrow \infty$. (This was also checked by
numerical calculations not shown here.) For a finite $T_{20}$, the
finite broadening of the exciton peak may destroy the linear Fano
effect and we see a symmetric peak (Fig. 1). For
self-assembled SQD, the Fano asymmetry is well expressed even in
the linear regime (Fig.~5). It is interesting to note that, in the
regime of symmetric peak (see e.g. Fig. 1), an exciton frequency
shift $\sim G_R\sim 1/R^6$. However, in the regime of nonlinear
Fano effect (Fano shapes in Figs. 2,3) the resonant frequency
shift $\sim 1/R^3$. Another interesting feature of the nonlinear
regime is that the absorption never vanishes, even in the limit
$T_{20}, \tau_0 \rightarrow \infty$.

Qualitatively, the nonlinear Fano effect can be explained as
follows. When the SQD is strongly driven ($\Om_{0}\gg 1/T_2,1/\tau$),
the absorption peak
becomes strongly suppressed, as in an atom \c{Yariv}. For the
MNP, we assume that the plasmon is not strongly excited. This is because of
the very short lifetime of the plasmon (of order of $10~fs$).
Simultaneously, the ac dipole moments of the MNP and SQD increase with
increasing intensity. In this situation, the interference
between two channels of plasmon excitation in the MNP (these
channels correspond to the first and second terms in the total
electric field in Eq.~3) increases and the peak asymmetry
is greatly enhanced. For example, the depth of the minimum in
the absorption curve becomes comparable to the peak hight in
Fig.~2.

It is possible to show that the problem of F\"orster-like
interaction between SQD and MNP is equivalent to the Fano problem
\c{fano}. To solve this problem, we can also use the density matrix
formulation for a description of the plasmon excitations in MNP and
look at the interaction between continuum plasmon states and
discrete exciton states directly, without employing a
self-consistent approach. Then by using the
fluctuation-dissipation theory \c{fludiss,Govorov_NL_2006}, we are
able to recover the previous results in the linear regime with
fast relaxation in MNP.

In conclusion, we have studied the optical properties of a hybrid
nanostructure composed of a MNP and a SQD. The interaction between
plasmon and exciton leads to interesting effects such as F\"orster
energy transfer, exciton energy shift, and interference. The
energy absorption and Rayleigh scattering reveal the formation of
collective hybrid excitons. At high light intensity, we find a
novel nonlinear Fano resonance which has striking differences to
the usual Fano effect.

{\bf Acknowledgments} We acknowledge helpful discussions
with Dr. T. Klar in Physics Department and CeNS, Ludwig-Maximiliams-Universität Münchën, Germany. 
This work was supported by NIST and BioTechnology Initiative at Ohio University.


\end{document}